\documentclass[12pt]{article}
\usepackage{amssymb}
\renewcommand{\theequation}{\arabic{section}.\arabic{equation}}

\begin{document}



\def\a{\alpha}
\def\b{\beta}
\def\d{\delta}
\def\e{\epsilon}
\def\g{\gamma}
\def\h{\mathfrak{h}}
\def\k{\kappa}
\def\l{\lambda}
\def\o{\omega}
\def\p{\wp}
\def\r{\rho}
\def\t{\theta}
\def\s{\sigma}
\def\z{\zeta}
\def\x{\xi}
 \def\A{{\cal{A}}}
 \def\B{{\cal{B}}}
 \def\C{{\cal{C}}}
 \def\D{{\cal{D}}}
\def\G{\Gamma}
\def\K{{\cal{K}}}
\def\O{\Omega}
\def\R{\bar{R}}
\def\T{{\cal{T}}}
\def\L{\Lambda}
\def\f{E_{\tau,\eta}(sl_2)}
\def\E{E_{\tau,\eta}(sl_n)}
\def\Zb{\mathbb{Z}}
\def\Cb{\mathbb{C}}

\def\R{\overline{R}}

\def\beq{\begin{equation}}
\def\eeq{\end{equation}}
\def\bea{\begin{eqnarray}}
\def\eea{\end{eqnarray}}
\def\ba{\begin{array}}
\def\ea{\end{array}}
\def\no{\nonumber}
\def\le{\langle}
\def\re{\rangle}
\def\lt{\left}
\def\rt{\right}

\newtheorem{Theorem}{Theorem}
\newtheorem{Definition}{Definition}
\newtheorem{Proposition}{Proposition}
\newtheorem{Lemma}{Lemma}
\newtheorem{Corollary}{Corollary}
\newcommand{\proof}[1]{{\bf Proof. }
        #1\begin{flushright}$\Box$\end{flushright}}

\baselineskip=20pt

\newfont{\elevenmib}{cmmib10 scaled\magstep1}
\newcommand{\preprint}{
   \begin{flushleft}
   \end{flushleft}\vspace{-1.3cm}
   \begin{flushright}\normalsize
     {\tt hep-th/0512154} \\ February 2006
   \end{flushright}}
\newcommand{\Title}[1]{{\baselineskip=26pt
   \begin{center} \Large \bf #1 \\ \ \\ \end{center}}}
\newcommand{\Author}{\begin{center}
   \large \bf
Wen-Li Yang$,{}^{a,b}$~ and~Yao-Zhong Zhang${}^b$\end{center}}
\newcommand{\Address}{\begin{center}

     ${}^a$ Institute of Modern Physics, Northwest University,
     Xian 710069, P.R. China\\
     ${}^b$ Department of Mathematics, University of Queensland, Brisbane, QLD 4072,
     Australia
   \end{center}}
\newcommand{\Accepted}[1]{\begin{center}
   {\large \sf #1}\\ \vspace{1mm}{\small \sf Accepted for Publication}
   \end{center}}

\preprint
\thispagestyle{empty}
\bigskip\bigskip\bigskip

\Title{$T$-$Q$ relation  and exact solution for the XYZ chain with
general nondiagonal boundary terms} \Author

\Address
\vspace{1cm}

\begin{abstract}
We propose that the  Baxter's $Q$-operator for the quantum XYZ
spin chain with open boundary conditions is given by the
$j\rightarrow \infty$ limit of the corresponding transfer matrix
with spin-$j$ (i.e., $(2j+1)$-dimensional) auxiliary space. The
associated $T$-$Q$ relation is derived from the fusion hierarchy
of the model. We use this relation to determine the Bethe Ansatz
solution of the eigenvalues of the fundamental transfer matrix.
The solution yields the complete spectrum of the Hamiltonian.

\vspace{1truecm} \noindent {\it PACS:} 75.10.Pq, 04.20.Jb,
05.50.+q

\noindent {\it Keywords}: Spin chain; reflection equation; Bethe
Ansatz; $Q$-operator; fusion hierarchy
\end{abstract}
\newpage
\section{Introduction}
\label{intro} \setcounter{equation}{0}

In this paper we are interested in the exact solution of the
quantum XYZ quantum spin chain with most general nondiagonal
boundary terms, defined by the Hamiltonian \bea
H&=&\sum^{N-1}_{j=1}\lt(J_x\s^x_j\s^x_{j+1}
+J_y\s^y_j\s^y_{j+1}+J_z\s^z_j\s^z_{j+1}\rt)\,+\, h^{(-)}_x\s^x_1
+h^{(-)}_y\s^y_1+h^{(-)}_z\s^z_1\no\\
&&\qquad + h^{(+)}_x\s^x_N
+h^{(+)}_y\s^y_N+h^{(+)}_z\s^z_N,\label{Ham}\eea where
$\s^x,\,\s^y,\,\s^z$ are the usual Pauli matrices, $N$ is the
number of spins, the bulk coupling constants $J_x,\,J_y,\,J_z$ are
related to the crossing parameter $\eta$ and modulus parameter
$\tau$ by the following relations, \bea
J_x=\frac{e^{i\pi\eta}\s(\eta+\frac{\tau}{2})}{\s(\frac{\tau}{2})},
\quad J_y=\frac{e^{i\pi\eta}\s(\eta+\frac{1}{2}+\frac{\tau}{2})}
{\s(\frac{1}{2}+\frac{\tau}{2})}, \quad
J_z=\frac{\s(\eta+\frac{1}{2})}{\s(\frac{1}{2})}, \no\eea and the
components of boundary magnetic fields  associated with the left
and right boundaries are given by \bea h^{(\mp)}_x&=&\pm
e^{-i\pi(\sum_{l=1}^3\a^{(\mp)}_l-\frac{\tau}{2})}\frac{\s(\eta)}
{\s(\frac{\tau}{2})}\prod_{l=1}^3\frac{\s(\a^{(\mp)}_l-\frac{\tau}{2})}
{\s(\a^{(\mp)}_l)},\quad h^{(\mp)}_z=\pm \frac{\s(\eta)}
{\s(\frac{1}{2})}\prod_{l=1}^3\frac{\s(\a^{(\mp)}_l-\frac{1}{2})}
{\s(\a^{(\mp)}_l)},\no\\
h^{(\mp)}_y&=&\pm
e^{-i\pi(\sum_{l=1}^3\a^{(\mp)}_l-\frac{1}{2}-\frac{\tau}{2})}\frac{\s(\eta)}
{\s(\frac{1}{2}+\frac{\tau}{2})}\prod_{l=1}^3
\frac{\s(\a^{(\mp)}_l-\frac{1}{2}-\frac{\tau}{2})}
{\s(\a^{(\mp)}_l)}. \label{Boundary}\eea Here $\s(u)$ is the
$\s$-function defined in the next section and $\{\a^{(\mp)}_l\}$
are free boundary parameters which specify the boundary coupling
(equivalently, the boundary magnetic fields).

In solving the closed XYZ chain (with periodic boundary
condition), Baxter constructed a $Q$-operator \cite{Bax71}, which
has now been proved to be a fundamental object in the theory of
exactly solvable models \cite{Bax82}. However, Baxter's original
construction of the $Q$-operator was ad hoc and its connection
with the quantum inverse scattering method was not clear. It was
later argued \cite{Baz97} that the $Q$-operator for the XXZ spin
chain with periodic boundary condition may be constructed from the
$U_q(\widehat{sl_2})$ universal L-operator whose associated
auxiliary space is taken  as an infinite dimensional q-oscillator
representation of $U_q(\widehat{sl_2})$. (See also \cite{Bax04}
for recent discussions on the direct construction of the
$Q$-operator of the closed chain.) In \cite{Yan06}, a natural
set-up to define the Baxter's $Q$-operator within the quantum
inverse scattering method was proposed for the open XXZ chain.
However, the generalization of the construction to {\it
off-critical\/} elliptic solvable models \cite{Pea87} (including
the XYZ chain) is still lacking even for the closed chains.

In this paper, we generalize the construction in \cite{Yan06} to
Baxter's eight-vertex model and propose that the $Q$-operator
$\bar{Q}(u)$ for the XYZ quantum spin chain (closed or open) is
given by the $j\rightarrow \infty$ limit of the corresponding
transfer matrix $t^{(j)}(u)$ with spin-$j$ (i.e.,
$(2j+1)$-dimensional) auxiliary space, \bea \bar{Q}(u) =
\lim_{j\rightarrow \infty}t^{(j)}(u-2j\eta). \label{Qbar} \eea
This relation together with the fusion hierarchy of the transfer
matrix leads to the $T$-$Q$ relation. We then use the $T$-$Q$
relation, together with some additional properties of the transfer
matrix, to determine the complete set of the eigenvalues  and
Bethe Ansatz equations of the transfer matrix associated with the
Hamiltonian (\ref{Ham}) under certain constraint of the boundary
parameters (see (\ref{Restr-1}) below).

The paper is organized as follows.  In Section 2, we introduce our
notation and some basic ingredients.  In Section 3, we derive the
$T$-$Q$ relation from (\ref{Qbar}) and the fusion hierarchy of the
open XXZ chain. In section 4, some properties of the fundamental
transfer matrix are obtained.  By means of these properties and
the $T$-$Q$ relation, we in Section 5 determine the eigenvalues of
the transfer matrix and the associated Bethe Ansatz equations,
thus giving the complete spectrum of the Hamiltonian (\ref{Ham})
under the constraint (\ref{Restr-1}) of the boundary parameters.
We summarize our conclusions in Section 6. Some detailed technical
calculations are given in Appendices A-C.


\section{ Fundamental transfer matrix}
\label{XXZ} \setcounter{equation}{0}

Let us fix $\tau$ such that ${\rm Im}(\tau)>0$ and a generic
complex number $\eta$. Introduce the following elliptic functions
\bea &&\t\lt[
\begin{array}{c}
a\\b
\end{array}\rt](u,\tau)=\sum_{m=-\infty}^{\infty}
\exp\lt\{i\pi\lt[(m+a)^2\tau+2(m+a)(u+b)\rt]\rt\},\label{Function-a-b}\\
&&\s(u)=\t\lt[\begin{array}{c}\frac{1}{2}\\[2pt]\frac{1}{2}
\end{array}\rt](u,\tau),\quad \zeta(u)=\frac{\partial}{\partial u}
\lt\{\ln\s(u)\rt\}.\label{Function} \eea Among them the
$\s$-function\footnote{Our $\s$-function is the
$\vartheta$-function $\vartheta_1(u)$ \cite{Whi50}. It has the
following relation with the {\it Weierstrassian\/} $\s$-function
if denoted it by $\s_w(u)$: $\s_w(u)\propto e^{\eta_1u^2}\s(u)$,
$\eta_1=\pi^2(\frac{1}{6}-4\sum_{n=1}^{\infty}\frac{nq^{2n}}{1-q^{2n}})
$ and $q=e^{i\tau}$.}
 satisfies the following
identity:\bea
&&\s(u+x)\s(u-x)\s(v+y)\s(v-y)-\s(u+y)\s(u-y)\s(v+x)\s(v-x)\no\\
&&~~~~~~=\s(u+v)\s(u-v)\s(x+y)\s(x-y),\label{identity}\eea which
will be useful in deriving equations in the following.

The well-known eight-vertex model R-matrix $R(u)\in {\rm
End}(\Cb^2\otimes \Cb^2)$ is given by  \bea
R(u)=\lt(\begin{array}{llll}a(u)&&&d(u)\\&b(u)&c(u)&\\
&c(u)&b(u)&\\d(u)&&&a(u)\end{array}\rt). \label{r-matrix}\eea The
non-vanishing matrix elements  are \cite{Bax82}\bea
&&\hspace{-0.8truecm}a(u)\hspace{-0.1truecm}=
\hspace{-0.1truecm}\frac{\t\lt[\begin{array}{c} 0\\\frac{1}{2}
\end{array}\rt]\hspace{-0.16truecm}(u,2\tau)\hspace{0.12truecm}
\t\lt[\begin{array}{c} \frac{1}{2}\\[2pt]\frac{1}{2}
\end{array}\rt]\hspace{-0.16truecm}(u+\eta,2\tau)}{\t\lt[\begin{array}{c} 0\\\frac{1}{2}
\end{array}\rt]\hspace{-0.16truecm}(0,2\tau)\hspace{0.12truecm}
\t\lt[\begin{array}{c} \frac{1}{2}\\[2pt]\frac{1}{2}
\end{array}\rt]\hspace{-0.16truecm}(\eta,2\tau)},\quad
b(u)\hspace{-0.1truecm}=\hspace{-0.1truecm}\frac{\t\lt[\begin{array}{c}
\frac{1}{2}\\[2pt]\frac{1}{2}
\end{array}\rt]\hspace{-0.16truecm}(u,2\tau)\hspace{0.12truecm}
 \t\lt[\begin{array}{c} 0\\\frac{1}{2}
\end{array}\rt]\hspace{-0.16truecm}(u+\eta,2\tau)}
{\t\lt[\begin{array}{c} 0\\\frac{1}{2}
\end{array}\rt]\hspace{-0.16truecm}(0,2\tau)\hspace{0.12truecm}
\t\lt[\begin{array}{c} \frac{1}{2}\\[2pt]\frac{1}{2}
\end{array}\rt]\hspace{-0.16truecm}(\eta,2\tau)},\no\\
&&\hspace{-0.8truecm}c(u)\hspace{-0.1truecm}=
\hspace{-0.1truecm}\frac{\t\lt[\begin{array}{c} 0\\\frac{1}{2}
\end{array}\rt]\hspace{-0.16truecm}(u,2\tau)\hspace{0.12truecm}
 \t\lt[\begin{array}{c} 0\\\frac{1}{2}
\end{array}\rt]\hspace{-0.16truecm}(u+\eta,2\tau)}
{\t\lt[\begin{array}{c} 0\\\frac{1}{2}
\end{array}\rt]\hspace{-0.16truecm}(0,2\tau)\hspace{0.12truecm}
\t\lt[\begin{array}{c} 0\\\frac{1}{2}
\end{array}\rt]\hspace{-0.16truecm}(\eta,2\tau)},\quad
d(u)\hspace{-0.1truecm}=\hspace{-0.1truecm}\frac{\t\lt[\begin{array}{c}
\frac{1}{2}\\[2pt]\frac{1}{2}
\end{array}\rt]\hspace{-0.16truecm}(u,2\tau)\hspace{0.12truecm}
 \t\lt[\begin{array}{c} \frac{1}{2}\\[2pt]\frac{1}{2}
\end{array}\rt]\hspace{-0.16truecm}(u+\eta,2\tau)}
{\t\lt[\begin{array}{c} 0\\\frac{1}{2}
\end{array}\rt]\hspace{-0.16truecm}(0,2\tau)\hspace{0.12truecm}
\t\lt[\begin{array}{c} 0\\\frac{1}{2}
\end{array}\rt]\hspace{-0.16truecm}(\eta,2\tau)}.\label{r-func}\eea
Here $u$ is the spectral parameter and $\eta$ is the so-called
crossing parameter. The R-matrix satisfies the quantum Yang-Baxter
equation \bea R_{1,2}(u_1-u_2)R_{1,3}(u_1-u_3)R_{2,3}(u_2-u_3)
=R_{2,3}(u_2-u_3)R_{1,3}(u_1-u_3)R_{1,2}(u_1-u_2),\label{QYB}\eea
and the properties, \bea
&&\hspace{-1.5cm}\mbox{PT-symmetry}:\hspace{0.68truecm}
R_{1,2}(u)=R_{2,1}(u)=R_{1,2}^{t_1t_2}(u),\label{PT}\\
&&\hspace{-1.5cm}\mbox{$Z_2$-symmetry}: \,
\qquad\s^i_1\s^i_2R_{1,2}(u)=R_{1,2}(u)\s^i_1\s^i_2,\quad
\mbox{for}\,\,
i=x,y,z,\label{Z2-sym}\\
 &&\hspace{-1.5cm}\mbox{Unitarity
relation}:\,R_{1,2}(u)R_{2,1}(-u)= -\xi(u)\,{\rm id},
\quad \xi(u)=\frac{\s(u+\eta)\s(u-\eta)}{\s(\eta)\s(\eta)},\label{Unitarity}\\
&&\hspace{-1.5cm}\mbox{Crossing
relation}:\,R_{1,2}(u)=V_1R_{1,2}^{t_2}(-u-\eta)V_1,\quad
V=-i\s^y. \label{crosing-unitarity}\eea
Here 
$R_{2,1}(u)=P_{12}R_{1,2}(u)P_{12}$ with $P_{12}$ being the usual
permutation operator and $t_i$ denotes transposition in the $i$-th
space. Throughout this paper we adopt the standard notations: for
any matrix $A\in {\rm End}(\Cb^2)$, $A_j$ is an embedding operator
in the tensor space $\Cb^2\otimes \Cb^2\otimes\cdots$, which acts
as $A$ on the $j$-th space and as identity on the other factor
spaces; $R_{i,j}(u)$ is an embedding operator of R-matrix in the
tensor space, which acts as identity on the factor spaces except
for the $i$-th and $j$-th ones.

Integrable open spin chains can be constructed as follows
\cite{Skl88}. Let us introduce a pair of K-matrices $K^-(u)$ and
$K^+(u)$. The former satisfies the reflection equation (RE)
\begin{eqnarray}
 &&R_{1,2}(u_1-u_2)K^-_1(u_1)R_{2,1}(u_1+u_2)K^-_2(u_2)\no\\
  &&~~~~~~=
 K^-_2(u_2)R_{1,2}(u_1+u_2)K^-_1(u_1)R_{2,1}(u_1-u_2),\label{RE-V}
\end{eqnarray}
and the latter  satisfies the dual RE \cite{Skl88,Che84}\bea
&&R_{1,2}(u_2-u_1)K^+_1(u_1)\,\,R_{2,1}(-u_1-u_2-2\eta)\,K^+_2(u_2)\no\\
&&\qquad\quad=
K^+_2(u_2)R_{1,2}(-u_1-u_2-2\eta)\,K^{+}_1(u_1)R_{2,1}(u_2-u_1).
\label{DRE-V1}\eea Then the transfer matrix $t(u)$ of the open XYZ
chain with general integrable boundary terms is given by
\cite{Skl88} \bea
t(u)=tr_0\lt(K_0^+(u)T_0(u)K^-_0(u)\hat{T}_0(u)\rt),\label{trans}\eea
where $T_0(u)$ and $\hat{T}_0(u)$ are the monodromy matrices \bea
T_0(u)=R_{0,N}(u)\ldots R_{0,1}(u),\quad
\hat{T}_0(u)=R_{1,0}(u)\ldots R_{N,0}(u),\label{Mon}\eea and
$tr_0$ denotes trace over the ``auxiliary space" $0$.

In this paper, we consider the most general solutions $K^{\mp}(u)$
\cite{Ina94} to the associated reflection equation and its dual,

\bea
K^{-}(u)&=&\hspace{-0.2cm}\frac{\s(2u)}{2\s(u)}\lt\{I\hspace{-0.1cm}
+\hspace{-0.1cm} \frac{c^{(-)}_x\s(u)e^{-i\pi
u}}{\s(u+\frac{\tau}{2})}\s^x\hspace{-0.1truecm}
+\hspace{-0.1cm}\frac{c^{(-)}_y\s(u)e^{-i\pi
u}}{\s(u+\frac{1+\tau}{2})}\s^y  \hspace{-0.1cm}+ \hspace{-0.1cm}
\frac{c^{(-)}_z\s(u)}{\s(u+\frac{1}{2})}\s^z
\rt\},\label{K-matrix1}\\
K^{+}(u)&=&\hspace{-0.2cm}\lt.K^-(-u-\eta)\rt|_{\{c^{(-)}_l\}\rightarrow
\{c^{(+)}_l\}},\label{K-matrix2} \eea where $I$ is the $2\times 2$
identity matrix and the constants $\{c^{(\mp)}_l\}$ are expressed
in terms of boundary parameters $\{\a^{(\mp)}_l\}$ as follows:
\bea c^{(\mp)}_x&=&e^{-i\pi(\sum_l\a^{(\mp)}_l-\frac{\tau}{2})}
\prod_{l=1}^3\frac{\s(\a^{(\mp)}_l-\frac{\tau}{2})}
{\s(\a^{(\mp)}_l)}, \quad  c^{(\mp)}_z=
\prod_{l=1}^3\frac{\s(\a^{(\mp)}_l-\frac{1}{2})}
{\s(\a^{(\mp)}_l)},\no\\
c^{(\mp)}_y&=&e^{-i\pi(\sum_l\a^{(\mp)}_l-\frac{1}{2}-\frac{\tau}{2})}
\prod_{l=1}^3\frac{\s(\a^{(\mp)}_l-\frac{1}{2}-\frac{\tau}{2})}
{\s(\a^{(\mp)}_l)}.\eea Sklyanin has shown that the transfer
matrices with different spectral parameters commute with each
other: $[t(u)\,, t(v)]=0$. This ensures the integrability of the
open XYZ chain. The Hamiltonian (\ref{Ham}) can be expressed in
terms of the transfer matrix \bea
H=\frac{\s(\eta)}{\s'(0)}\lt\{\lt.\frac{\partial}{\partial u}\,\ln
t(u)\rt|_{u=0}-\lt[(N-1)\zeta(\eta)+2\zeta(2\eta)\rt]\rt\},
\label{Ham-rel}\eea where $\s'(0)=\lt.\frac{\partial}{\partial
u}\,\s(u)\rt|_{u=0}$.


\section{ Fusion hierarchy and $T$-$Q$ relation}
\label{T-QR} \setcounter{equation}{0}

We shall use the fusion procedure, which was first developed for
R-matrices \cite{Kar79} and then later generalized  for K-matrices
\cite{Mez92,Zho96},  to obtain the Baxter $T$-$Q$ relation. The
fused spin-$(j,\frac{1}{2})$ R-matrix
$(j=\frac{1}{2},1,\frac{3}{2},\ldots)$ is given by \cite{Kar79}
\bea R_{\langle1\ldots 2j\rangle,2j+1}(u)=P^{(+)}_{1\ldots2j}
R_{1,2j+1}(u)R_{2,2j+1}(u+\eta)\cdots R_{2j,2j+1}(u+(2j-1)\eta)
P^{(+)}_{1\ldots2j},\eea where $P^{(+)}_{1\ldots2j}$ is the
completely symmetric projector.
Following \cite{Mez92,Zho96}, the fused spin-$j$ K-matrix
$K^-_{\langle1\ldots 2j\rangle}(u)$ is given by  \bea
K^-_{\langle1\ldots
2j\rangle}(u)&=&P^{(+)}_{1\ldots2j}\,\lt\{K^-_{2j}(u)R_{2j,2j-1}(2u+\eta)
K^-_{2j-1}(u+\eta)\rt.\no\\
&&\times
R_{2j,2j-2}(2u+2\eta)R_{2j-1,2j-2}(2u+3\eta)K^-_{2j-2}(u+2\eta)\cdots
\no\\
&&\times \lt.R_{2,1}(2u+(4j-3)\eta)
K^-_1(u+(2j-1)\eta)\rt\}\,P^{(+)}_{1\ldots2j}. \eea The fused
spin-$j$ K-matrix $K^+_{\langle1\ldots 2j\rangle}(u)$ is given by
\bea K^+_{\langle1\ldots 2j\rangle}(u)
=F(u|2j)\,\lt.K^-_{\langle1\ldots
2j\rangle}(-u-2j\eta)\rt|_{\{c^{(-)}_l\}\rightarrow
\{c^{(+)}_l\}}, \eea where the scalar functions $F(u|q)$ are given
by $
F(u|q)={1/{(\prod_{l=1}^{q-1}\prod_{k=1}^{l}\xi(2u+l\eta+k\eta))}}$,
for $q=1,2,\ldots$. The fused transfer matrix $t^{(j)}(u)$
constructed with a spin-$j$ auxiliary space is given by \bea
t^{(j)}(u)=tr_{1\ldots 2j}\lt(K^+_{\langle1\ldots
2j\rangle}(u)T_{\langle1\ldots 2j\rangle}(u)K^-_{\langle1\ldots
2j\rangle}(u)\hat{T}_{\langle1\ldots 2j\rangle}(u+(2j-1)\eta)
\rt),\, j=\frac{1}{2},1,\ldots,\eea where \bea T_{\langle1\ldots
2j\rangle}(u)&=&R_{\langle 1\ldots 2j\rangle,N}(u)\ldots
R_{\langle 1\ldots
2j\rangle,1}(u),\no\\
\hat{T}_{\langle1\ldots 2j\rangle}(u+(2j-1)\eta)&=&R_{\langle
1\ldots 2j\rangle,1}(u)\ldots R_{\langle 1\ldots 2j\rangle,N}(u).
\eea The transfer matrix (\ref{trans}) corresponds to the
fundamental case $j=\frac{1}{2}$, i.e., $t(u)=t^{(\frac{1}{2})}(u)
$. The fused transfer matrices constitute commutative families,
namely, \bea \lt[t^{(j)}(u),\,t^{(k)}(v)\rt]=0. \label{Com-1-1}
\eea They satisfy a so-called fusion hierarchy relation
\cite{Mez92,Zho96}\bea
t^{(j)}(u\hspace{-0.1truecm}-\hspace{-0.1truecm}(2j-1)\eta)
\hspace{-0.1truecm}=\hspace{-0.1truecm}
t^{(j-\frac{1}{2})}(u\hspace{-0.1truecm}-\hspace{-0.1truecm}(2j-1)\eta)
t(u)-\d(u)t^{(j-1)}(u\hspace{-0.1truecm}-\hspace{-0.1truecm}(2j-1)\eta),
\,j=1,\frac{3}{2},\ldots.\label{fusion-1}\eea In the above
hierarchy, we have used the convention $t^{(0)}(u)={\rm id}$. The
coefficient function $\d(u)$ can be expressed in terms of the
quantum determinants of the monodromy matrices $T(u)$,
$\hat{T}(u)$ and K-matrices \cite{Mez92,Zho96}, \bea
\d(u)=\frac{\rm{Det}\lt\{T(u)\rt\}\,\rm{Det}\lt\{\hat{T}(u)\rt\}\,
\rm{Det}\lt\{K^-(u)\rt\}
\,\rm{Det}\lt\{K^+(u)\rt\}}{\tilde{\rho}_{1,1}(2u-\eta)}.\label{D-1}\eea
Here
$\tilde{\rho}_{1,1}(u)=\frac{\s(-u)\s(u+2\eta)}{\s(\eta)\s(\eta)}$
 and the corresponding determinants are given by \cite{Zho96}
 \bea
\rm{Det}\lt\{T(u)\rt\}\,{\rm
id}&=&tr_{12}\lt(P^{(-)}_{12}T_1(u-\eta)T_2(u)P^{(-)}_{12}\rt)
=\lt(\frac{\s(u+\eta)\s(u-\eta)}{\s(\eta)\s(\eta)}\rt)^{N}\,{\rm
id},\label{D-2}\\
\rm{Det}\lt\{\hat{T}(u)\rt\}\,{\rm
id}&=&tr_{12}\lt(P^{(-)}_{12}T_1(u-\eta)T_2(u)P^{(-)}_{12}\rt)
=\lt(\frac{\s(u+\eta)\s(u-\eta)}{\s(\eta)\s(\eta)}\rt)^{N}\,{\rm
id},\\
\rm{Det}\lt\{K^-(u)\rt\}&=&tr_{12}
\lt(P^{(-)}_{12}K^-_1(u-\eta)R_{12}(2u-\eta)K^-_2(u)\rt),\label{Delta-1}\\
\rm{Det}\lt\{K^+(u)\rt\}&=&tr_{12}
\lt(P^{(-)}_{12}K^+_2(u)R_{12}(-2u-\eta)K^+_1(u-\eta)\rt),\label{Delta-2}
 \eea where $P^{(-)}_{12}$ is the completely antisymmetric
 project: $P^{(-)}_{12}=\frac{1}{2}\lt({\rm id}-P_{12}\rt)$. Using
 the crossing relations of the K-matrices (see
 (\ref{K-cross-1})-(\ref{K-cross-4}) below) and after a tedious calculation
 (for details see Appendix A),
 we find that the quantum determinants of the most general K-matrices
$K^{\pm}(u)$ given by (\ref{K-matrix1}) and (\ref{K-matrix2})
respectively are \bea
\rm{Det}\lt\{K^-(u)\rt\}&=&\frac{\s(2u-2\eta)}{\s(\eta)}\prod_{l=1}^{3}
\frac{\s(\a^{(-)}_l+u)\s(\a^{(-)}_l-u)}{\s(\a^{(-)}_l)\s(\a^{(-)}_l)},\label{D-2-1}\\
\rm{Det}\lt\{K^+(u)\rt\}&=&\frac{\s(2u+2\eta)}{\s(\eta)}\prod_{l=1}^{3}
\frac{\s(u+\a^{(+)}_l)\s(u-\a^{(+)}_l)}{\s(\a^{(+)}_l)\s(\a^{(+)}_l)}.\label{D-3}\eea
Substituting (\ref{D-2})-(\ref{D-3}) into (\ref{D-1}), one has
that \bea
&&\d(u)=\lt\{\frac{\s(u+\eta)\s(u-\eta)}{\s(\eta)\s(\eta)}\rt\}^{2N}
\frac{\s(2u-2\eta)\s(2u+2\eta)}{\s(2u-\eta)\s(2u+\eta)}\no\\
&&\qquad\qquad\times\prod_{\g=\pm}\prod_{l=1}^3\frac{\s(u+\a^{(\g)}_l)\s(u-\a^{(\g)}_l)}{
\s(\a^{(\g)}_l)\s(\a^{(\g)}_l)}.\label{Coe-function} \eea

For generic $\eta$, the fusion hierarchy does not truncate (c.f.
the roots of unity case  in the trigonometric limit
\cite{Baz96,Nep04}). Hence $\{t^{(j)}(u)\}$ constitute an infinite
hierarchy, namely, $j$ taking values
$\frac{1}{2},1,\frac{3}{2},\ldots$. The commutativity
(\ref{Com-1-1}) of the fused transfer matrices $\{t^{(j)}(u)\}$
and the fusion relation (\ref{fusion-1}) imply that the
corresponding eigenvalue of the transfer matrix $t^{(j)}(u)$,
denoted by $\L^{(j)}(u)$, satisfies the following hierarchy
relation \bea
\L^{(j)}(u+\eta-2j\eta)&=&\L^{(j-\frac{1}{2})}(u-2(j-\frac{1}{2})\eta)\,
\L(u) -\d(u) \, \L^{(j-1)}(u-\eta-2(j-1)\eta),\no\\
&&\quad j=1,\frac{3}{2},\ldots. \label{fusion-2}\eea Here we have
used the convention $\L(u)=\L^{(\frac{1}{2})}(u)$ and
$\L^{(0)}(u)=1$. Dividing both sides of (\ref{fusion-2}) by
$\L^{(j-\frac{1}{2})}(u-2(j-\frac{1}{2})\eta)$, we have \bea
\L(u)=\frac{\L^{(j)}(u+\eta-2j\eta)}
{\L^{(j-\frac{1}{2})}(u-2(j-\frac{1}{2})\eta)}+\d(u)\,
\frac{\L^{(j-1)}(u-\eta-2(j-1)\eta)}
{\L^{(j-\frac{1}{2})}(u-2(j-\frac{1}{2})\eta)}.\label{fusion-3}\eea

We now consider the limit $j \rightarrow \infty$. We make the
fundamental assumption (\ref{Qbar}) (in particular,  that the
limit exists), which implies for the corresponding eigenvalues
\bea \bar{Q}(u)=\lim_{j\longrightarrow +\infty}\L^{(j)}(u-2j\eta)
, \eea where we have used the same notation for the operator
$\bar{Q}$ and its eigenvalues (c.f. Eq.(\ref{Qbar})). It follows
from (\ref{fusion-3}) that \bea
\L(u)=\frac{\bar{Q}(u+\eta)}{\bar{Q}(u)}+
\d(u)\,\frac{\bar{Q}(u-\eta)}{\bar{Q}(u)}.\label{fusion-4}\eea
Assuming the function $\bar{Q}(u)$ has the decomposition
$\bar{Q}(u)=f(u)Q(u)$ with \bea
Q(u)=\prod_{j=1}^{M}\s(u-u_j)\s(u+u_j+\eta), \label{fusion-5} \eea
where $M$ is certain non-negative integer and $\{u_j\}$ are some
parameters which will be specified later (see  (\ref{Eign-fuc})
and (\ref{BAE-2}) below), then Eq. (\ref{fusion-4}) becomes \bea
\L(u)= H_1(u)\,\frac{Q(u+\eta)}{Q(u)}+H_2(u)\,
\frac{Q(u-\eta)}{Q(u)}.\label{T-Q}\eea Here
$H_1(u)=\frac{f(u+\eta)}{f(u)}$ and
$H_2(u)=\d(u)\,\frac{f(u-\eta)}{f(u)}$. It is easy to see that the
functions $\{H_i(u)|i=1,2\}$ satisfy the relation \bea
H_1(u-\eta)H_2(u)=\d(u) \label{T-Q-1} \eea with $\d(u)$ given by
(\ref{Coe-function}).

In summary, the eigenvalue $\L(u)$ of the fundamental transfer
matrix $t(u)$ (\ref{trans}) has the decomposition form
(\ref{T-Q}), where the coefficient functions $\{H_i(u)\}$ satisfy
the constraint (\ref{T-Q-1}).  In the following, we shall use
certain properties of the eigenvalue $\L(u)$ derived from the
transfer matrix to determine the functions $\{H_i(u)\}$ and
therefore the eigenvalue $\L(u)$.

\section{Properties of the fundamental transfer matrix}
\label{Pro} \setcounter{equation}{0} Here we will derive five
properties of the fundamental transfer matrix $t(u)$, which
together with the $T$-$Q$ relation (\ref{T-Q})  enable us to
determine the functions $\{H_i(u)\}$.

In addition to the Riemann identity (\ref{identity}), the
$\s$-function enjoys the following identity:\bea
\s(2u)&=&\frac{2\s(u)\s(u+\frac{1}{2})\s(u+\frac{\tau}{2})
\s(u-\frac{1}{2}-\frac{\tau}{2})}{\s(\frac{1}{2})\s(\frac{\tau}{2})
\s(-\frac{1}{2}-\frac{\tau}{2})},\label{func-2}\\
\s(u+1)&=&-\s(u),\quad
\s(u+\tau)=-e^{-2i\pi(u+\frac{\tau}{2})}\s(u).\label{quasi-func}\eea
Moreover, from the definition of the elliptic function
(\ref{Function-a-b}), one may show that \bea \t\lt[
\begin{array}{c}
0\\\frac{1}{2}
\end{array}\rt](u+\tau,2\tau)
&=&e^{-i\pi(u+\frac{1}{2}+\frac{\tau}{2})}\,\,
\t\lt[\begin{array}{c} \frac{1}{2}\\[2pt]\frac{1}{2}
\end{array}\rt](u,2\tau),\label{func-3}\\
\t\lt[\begin{array}{c} \frac{1}{2}\\[2pt]\frac{1}{2}
\end{array}\rt](u+\tau,2\tau)
&=&e^{-i\pi(u+\frac{1}{2}+\frac{\tau}{2})}\,\,
\t\lt[\begin{array}{c}
0\\\frac{1}{2}\end{array}\rt](u,2\tau).\label{func-4}\eea

The property (\ref{func-2}) of the $\s$-function implies that the
matrix elements of the K-matrices $K^{\mp}(u)$ given by
(\ref{K-matrix1})-(\ref{K-matrix2}) are analytic functions of $u$.
Thus the commutativity of the transfer matrix $t(u)$ and the
analyticity of the R-matrix and K-matrices lead to  the following
analytic property: \bea  \mbox{ The eigenvalue of } t(u) \mbox{ is
an analytic function of $u$ at finite $u$}.\label{trans-Anal}\eea
The quasi-periodicity of the elliptic functions,
(\ref{quasi-func})-(\ref{func-4}), allows one to derive the
following properties of the R-matrix and K-matrices:\bea
R_{1,2}(u+1)&=&-\s^z_1R_{1,2}(u)\s^z_1=-\s^z_2R_{1,2}(u)\s^z_2,\quad
K^{\mp}(u+1)=-\s^zK^{\mp}(u)\s^z,\no\\
R_{1,2}(u+\tau)&=&-e^{-2i\pi(u+\frac{\eta}{2}+\frac{\tau}{2})}
\s^x_1R_{1,2}(u)\s^x_1=-e^{-2i\pi(u+\frac{\eta}{2}+\frac{\tau}{2})}
\s^x_2R_{1,2}(u)\s^x_2,\no\\
K^-(u+\tau)&=&-e^{-2i\pi(3u+\frac{3}{2}\tau)}\s^xK^-(u)\s^x,\no\\
K^+(u+\tau)&=&-e^{-2i\pi(3u+3\eta+\frac{3}{2}\tau)}\s^xK^+(u)\s^x.\no\eea
From these relations one obtains the quasi-periodic properties of
$t(u)$, \bea t(u+1)=t(u),\quad
t(u+\tau)=e^{-2i\pi(N+3)(2u+\eta+\tau)}\,t(u).\label{trans-Perio}\eea
The initial conditions of the R-matrix $R_{12}(0)=P_{12}$ and
K-matrices: $K^-(0)={\rm id}$, $trK^+(0)={\s(2\eta)}/{\s(\eta)}$,
imply that the initial condition of the fundamental transfer
matrix $t(u)$ is given by \bea
\lt.t(u)\rt|_{u=0}=\frac{\s(2\eta)}{\s(\eta)}\times {\rm
id}.\label{trans-initial}\eea The identity \bea
\frac{\s(u)}{\s(\frac{\tau}{2})}=\frac{ \theta\lt[\begin{array}{l}
0\\\frac{1}{2}\end{array}\rt](u,2\tau)\,\,\theta\lt[\begin{array}{l}
\frac{1}{2}\\[2pt]\frac{1}{2}\end{array}\rt](u,2\tau)}
{\theta\lt[\begin{array}{l}
0\\\frac{1}{2}\end{array}\rt](\frac{\tau}{2},2\tau)\,\,
\theta\lt[\begin{array}{l}
\frac{1}{2}\\[2pt]\frac{1}{2}\end{array}\rt]
(\frac{\tau}{2},2\tau)},\no\eea suggests the semi-classical
property of the R-matrix $\lim_{\eta\rightarrow
0}\lt\{\s(\eta)R_{12}(u)\rt\}=\s(u)\,{\rm id}$. This enables us to
derive  the semi-classical property of $t(u)$, \bea
\hspace{-0.2cm}\lim_{\eta\rightarrow 0}\hspace{-0.1cm}\lt\{
\s^{2N}(\eta)t(u)\rt\} \hspace{-0.1cm}=\hspace{-0.1cm}\s^{2N}(u)
\lim_{\eta\rightarrow
0}\hspace{-0.1cm}\lt\{tr(K^+(u)K^-(u))\rt\}\,\times{\rm
id}.\label{trans-semi}\eea

Now, we derive the crossing relation of $t(u)$. For this purpose,
we introduce \bea
\bar{K}_1^-(u)&=&tr_2\lt(P_{12}R_{12}(-2u-2\eta)\{K^-_2(u)\}^{t_2}\rt),
\label{K-cross-1}\\
\bar{K}^+_1(u)&=&tr_2\lt(P_{12}R_{12}(2u)\{K^+_2(u)\}^{t_2}\rt).
\label{K-cross-2}\eea Through a straightforward calculation (see
Appendix A for details), we find \bea
\bar{K}^-(u)&=&-\frac{\s(2u)}{\s(\eta)}\s^yK^-(-u-\eta)\s^y
\equiv f_-(u)VK^-(-u-\eta)V,\label{K-cross-3}\\
\bar{K}^+(u)&=&\frac{\s(2u+2\eta)}{\s(\eta)}\s^yK^+(-u-\eta)\s^y
\equiv f_+(u)VK^+(-u-\eta)V,\label{K-cross-4}\eea where the
$2\times 2$ matrix $V$ is given in (\ref{crosing-unitarity}),
$f_-(u)=\frac{\s(2u)}{\s(\eta)}$ and
$f_+(u)=-\frac{\s(2u+2\eta)}{\s(\eta)}$. The PT-symmetry
(\ref{PT}) and crossing relation (\ref{crosing-unitarity}) of the
R-matrix implies
\bea T_0^{t_0}(-u-\eta)&=&\lt(R_{0,N}(-u-\eta)\ldots
R_{0,1}(-u-\eta)\rt)^{t_0}\no\\
&\stackrel{(\ref{crosing-unitarity})}{=}&(-1)^{N-1}
\lt(V_0\,R_{0,N}^{t_N}(u)\ldots
R_{0,1}^{t_1}(u)\,V_0\rt)^{t_0}\no\\
&=&(-1)^{N-1} \lt(V_0\lt\{R_{0,N}(u)\ldots
R_{0,1}(u)\rt\}^{t_1\ldots t_N}V_0\rt)^{t_0}\no\\
&=&(-1)^{N-1}\,V_0\lt\{R_{0,N}(u)\ldots
R_{0,1}(u)\rt\}^{t_0\,t_1\ldots t_N}V_0\no\\
&\stackrel{(\ref{PT})}{=}&(-1)^{N-1} \,V_0 R_{0,1}(u)\ldots
R_{0,N}(u)\,V_0\stackrel{(\ref{PT})}{=}
(-1)^{N-1}V_0\hat{T}_0(u)V_0,\label{Dual-1}\eea Similarly, we have
another ``duality relations" between the monodromoy matrices
$T(u)$ and $\hat{T}(u)$,\bea
\hat{T}^{t_0}_0(-u-\eta)=(-1)^{N-1}V_0\,T_0(u)\,V_0.\label{Dual-2}\eea
The crossing relations of the K-matrices (\ref{K-cross-3}) and
(\ref{K-cross-4}), and the ``duality relations" (\ref{Dual-1}) and
(\ref{Dual-2}), imply that  \bea
t(-u-\eta)&=&tr_0\lt(K^+_0(-u-\eta)T_0(-u-\eta)K^-_0(-u-\eta)
\hat{T}_0(-u-\eta)\rt)\no\\
&=&tr_0\lt(\lt\{K_0^+(-u-\eta)T_0(-u-\eta)\rt\}^{t_0}\lt\{K^-_0(-u-\eta)
\hat{T}_0(-u-\eta)\rt\}^{t_0}\rt)\no\\
&=&tr_0\lt(T_0^{t_0}(-u-\eta)(K_0^+(-u-\eta))^{t_0}
\hat{T}_0^{t_0}(-u-\eta)(K^-_0(-u-\eta))^{t_0}\rt)\no\\
&\stackrel{(\ref{Dual-1}),(\ref{Dual-2})}{=}&tr_0\lt(
\hat{T}_0(u)\lt\{V_0(K_0^+(-u-\eta))^{t_0}V_0\rt\}T_0(u)
\lt\{V_0(K_0^-(-u-\eta))^{t_0}V_0\rt\}\rt)\no\\
&\stackrel{(\ref{K-cross-3}),(\ref{K-cross-4})}{=}&
tr_0\lt(\hat{T}_0(u)(\bar{K}^+_0(u))^{t_0}T_0(u)
(\bar{K}^-_0(u))^{t_0}\rt)/f_+(u)f_-(u)\no\\
&\stackrel{(\ref{K-cross-1}),(\ref{K-cross-2})}{=}&
\hspace{-0.22truecm}\frac{tr_0tr_1tr_2\hspace{-0.1truecm}
\lt(\hspace{-0.1truecm}\hat{T}_0(u)P_{01}R_{0,1}(2u)K^+_1(u)
T_0(u)P_{02}R_{0,2}(-2u-2\eta)K_2^-(u)\hspace{-0.1truecm}\rt)}
{f_+(u)f_-(u)}
\no\\
&=& \hspace{-0.86truecm}\frac{tr_0tr_1tr_2\hspace{-0.1truecm}
\lt(\hspace{-0.1truecm}P_{01}\hat{T}_1(u)R_{0,1}(2u)T_0(u)
P_{02}R_{0,2}(-2u
\hspace{-0.1truecm}-\hspace{-0.1truecm}2\eta)K_2^-(u)
K^+_1(u)\hspace{-0.1truecm}\rt)}{f_+(u)f_-(u)}\no\\
&=& \hspace{-0.86truecm}\frac{tr_0tr_1tr_2\hspace{-0.1truecm}
\lt(\hspace{-0.1truecm}K^+_1(u)T_1(u)P_{01}R_{0,1}(2u)
P_{02}R_{0,2}(-2u
\hspace{-0.1truecm}-\hspace{-0.1truecm}2\eta)K_2^-(u)\hat{T}_1(u)
\hspace{-0.1truecm}\rt)}{f_+(u)f_-(u)}.\label{t-cross-1}\eea In
deriving the  second last equality of the above equation, we have
used the relation,
$\hat{T}_1(u)R_{0,1}(2u)T_0(u)=T_0(u)R_{0,1}(2u)\hat{T}_1(u)$,
which is a simple consequence of the quantum Yang-Baxter equation
(\ref{QYB}). Then, let us consider \bea
&&\hspace{-2truecm}tr_0tr_2\lt(P_{01}R_{0,1}(2u)P_{02}
R_{0,2}(-2u-2\eta)K^-_2(u)\rt)/f_+(u)f_-(u)\no\\
&=& tr_0\lt(P_{01}R_{0,1}(2u)tr_2\lt\{P_{02}
R_{0,2}(-2u-2\eta)K^-_2(u)\rt\}\rt)/f_+(u)f_-(u)\no\\
&\stackrel{(\ref{K-cross-1})}{=}& tr_0\lt(P_{01}R_{0,1}(2u)
\lt\{\bar{K}^-_0(u)\rt\}^{t_0}\rt)/f_+(u)f_-(u)\no\\
&\stackrel{(\ref{K-cross-3})}{=}& tr_0\lt(V_0P_{01}R_{0,1}(2u)V_0
\lt\{K^-_0(-u-\eta)\rt\}^{t_0}\rt)/f_+(u)\no\\
&\stackrel{(\ref{Z2-sym})}{=}& tr_0\lt(V_1P_{01}R_{0,1}(2u)V_1
\lt\{K^-_0(-u-\eta)\rt\}^{t_0}\rt)/f_+(u)\no\\
&\stackrel{(\ref{K-cross-1})}{=}& V_1
\bar{K}^-_1(-u-\eta)V_1/f_+(u) \stackrel{(\ref{K-cross-3})}{=}
\frac{f_-(-u-\eta)}{f+(u)}K_1^-(u)=K_1^-(u)\no\eea Substituting
the above equation into (\ref{t-cross-1}), we have the following
crossing relation of $t(u)$:\bea
t(-u-\eta)=tr_1\lt(K^+_1(u)T_1(u)K^-_1(u)\hat{T}_1(u)
\rt)=t(u).\label{trans-cross}\eea We remark that our above proof
of the crossing relation (\ref{trans-cross}) generalizes the
previous proofs given in \cite{Beh96}.


\section{ Eigenvalues and Bethe Ansatz equations}
\label{BAE} \setcounter{equation}{0}

It follows from (\ref{trans-Anal})-(\ref{trans-semi}) and
(\ref{trans-cross}) that the eigenvalue $\L(u)$, as a function of
$u$, has the following properties, \bea
&&\hspace{-1.5cm}\mbox{Analyticity}:\,\qquad\quad \L(u) \mbox{ is
an
analytic function of $u$ at finite $u$}.\label{Eigen-Anal}\\
&&\hspace{-1.5cm}\mbox{Quasi-Periodicity}:\,\qquad\quad
\L(u+1)=\L(u),\quad
\L(u+\tau)=e^{-2i\pi(N+3)(2u+\eta+\tau)}\L(u), \label{Eign-Per}\\
&&\hspace{-1.5cm}\mbox{Crossing
symmetry}:\,\quad\quad\L(-u-\eta)=\L(u),\label{Eigen-Cro}\\
&&\hspace{-1.5cm}\mbox{Initial
condition}:\,\qquad\qquad\L(0)=\frac{\s(2\eta)}{\s(\eta)},\label{Eigen-In}\\
&&\hspace{-1.5cm}\mbox{Semi-classic property}:\,\,\quad
\lim_{\eta\rightarrow 0}\hspace{-0.1cm}\lt\{
\s^{2N}(\eta)\L(u)\rt\} \hspace{-0.1cm}=\hspace{-0.1cm}\s^{2N}(u)
\lim_{\eta\rightarrow
0}\hspace{-0.1cm}\lt\{tr(K^+(u)K^-(u))\rt\}.\label{Eigen-semi}
 \eea The $T$-$Q$ relations (\ref{T-Q}) and (\ref{T-Q-1}), together with
the above properties (\ref{Eigen-Anal})-(\ref{Eigen-semi}),  can
be used to determine $\{H_i(u)\}$ and the eigenvalues of the
fundamental transfer matrix.

For this purpose, we define 6 {\it discrete\/} parameters
$\{\e^{(\g)}_l=\pm 1|\g=\pm,\,l=1,2,3\}$ and fix $\e^{(-)}_1=+1$
in the following. Associated with $\{\e^{(\g)}_l\}$, we introduce
\bea H^{(\pm)}_1(u|\{\e^{(\g)}_l\}) &=&\frac{\s^{2N}(u)\s(2u)}
{\s^{2N}(\eta)\s(2u+\eta)}\prod_{\g=\pm} \prod_{l=1}^3
\frac{\s(u+\eta\pm\e^{(\g)}_l\a_l^{(\g)})}
{\s(\e^{(\g)}_l\a_l^{(\g)})},\label{H-1}\\
H^{(\pm)}_2(u|\{\e^{(\g)}_l\})&=&
\frac{\s^{2N}(u+\eta)\s(2u+2\eta)}
{\s^{2N}(\eta)\s(2u+\eta)}\prod_{\g=\pm} \prod_{l=1}^3
\frac{\s(u\mp\e^{(\g)}_l\a_l^{(\g)})}
{\s(\e^{(\g)}_l\a_l^{(\g)})}.\label{H-2} \eea One may readily
check that the functions $H^{(\pm)}_1(u|\{\e^{(\g)}_l\})$ and
$H^{(\pm)}_2(u|\{\e^{(\g)}_l\})$ indeed satisfy (\ref{T-Q-1}),
namely, \bea
 H^{(\pm)}_1(u-\eta|\{\e^{(\g)}_l\})H^{(\pm)}_2(u|\{\e^{(\g)}_l\})
=\d(u). \eea The general solution to (\ref{T-Q-1}) can be written
as follows: \bea
H_1(u)=H^{(\pm)}_1(u|\{\e^{(\g)}_l\})\,g_1(u),\quad
H_2(u)=H^{(\pm)}_2(u|\{\e^{(\g)}_l\})\,g_2(u),\eea where
$\{g_i(u)\}$ satisfy the following relations, \bea
g_1(u-\eta)g_2(u)=1,\quad g_1(u+1)=g_1(u),\quad
g_2(u+1)=g_2(u).\label{Eq}\eea The solutions to (\ref{Eq}) have
the following form, \bea
g_1(u)=\l\frac{\prod_{j=1}^{N_2}\s(u-u^+_j)}
{\prod_{j=1}^{N_1}\s(u-u^-_j)},\quad
g_2(u)=\frac{1}{\l}\frac{\prod_{j=1}^{N_1}\s(u-u^-_j-\eta)}
{\prod_{j=1}^{N_2}\s(u-u^+_j-\eta)},\eea where $N_1$ and $N_2$ are
integers such that $N_1,\,N_2\geq 0$, and $\l$ is an non-zero
constant. In the above equation, we assume that $u^-_j\neq
\mp\e^{(\g)}_l\a^{(\g)}_l-\eta$ and $u^+_j\neq
\mp\e^{(\g')}_{l'}\a^{(\g')}_{l'}-\eta$, otherwise the
corresponding factors in $g_i(u)$  make transitions among
$\{H^{(\pm)}_1(u|\{\e^{(\g)}_l\})\}$
($\{H^{(\pm)}_2(u|\{\e^{(\g)}_l\})\}$ respectively). Then the
analyticity of $\L(u)$ (\ref{Eigen-Anal}) requires that $g_{1}(u)$
and $g_{2}(u)$ have common poles; i.e., $N_1=N_2$, and
$u^-_j=u^+_{j'}+\eta$. This means \bea
g_1(u)=\l\prod_{j=1}^{N_1}\frac{\s(u-u^-_j+\eta)}
{\s(u-u^-_j)},\quad
g_2(u)=\frac{1}{\l}\prod_{j=1}^{N_1}\frac{\s(u-u^-_j-\eta)}
{\s(u-u^-_j)}.\eea Since
$H^{(\pm)}_2(u|\{\e^{(\g)}_l\})=H^{(\pm)}_1(-u-\eta|\{\e^{(\g)}_l\})$,
the crossing symmetry of $\L(u)$ (\ref{Eigen-Cro}) implies that
$H_2(u)=H_1(-u-\eta)$. Hence, $N_1$ is even and \bea
g_1(u)&=&\l\prod_{j=1}^{\frac{N_1}{2}}\frac{\s(u-u^-_j+\eta)\s(u+u_j^-+2\eta)}
{\s(u-u^-_j)\s(u+u^-_j+\eta)},\\
g_2(u)&=&\l\prod_{j=1}^{\frac{N_1}{2}}\frac{\s(u-u^-_j-\eta)\s(u+u_j^-)}
{\s(u-u^-_j)\s(u+u^-_j+\eta)},\quad \l=\pm 1.\eea This is
equivalent to having additional Bethe roots; and the corresponding
factors, except $\l$, can be absorbed into those of $Q(u)$
(\ref{fusion-5}). Moreover the initial condition (\ref{Eigen-In})
implies $\l=+1$. Therefore the eigenvalues of the transfer matrix
take the following forms: \bea
H^{(\pm)}_1(u|\{\e^{(\g)}_l\})\,\frac{Q(u+\eta)}{Q(u)}
+H^{(\pm)}_2(u|\{\e^{(\g)}_l\})\,\frac{Q(u-\eta)}{Q(u)}.
\label{T-Q-3}\eea In the above expression, we still have an
unknown non-negative integer $M$ defined in (\ref{fusion-5}) to be
fixed.  Moreover, $\Lambda(u)$ must fulfill the properties
(\ref{Eign-Per}) and (\ref{Eigen-semi}). This provides the
following constraint to the boundary parameters $\{\a^{(\g)}_l\}$:
\bea
\sum_{\g=\pm}\sum_{l=1}^3\e^{(\g)}_l\a^{(\g)}_l=k\eta\quad{\rm
mod}(1),\qquad
\prod_{\g=\pm}\prod_{l=1}^3\e^{(\g)}_l=-1,\label{Restr-1}\eea
where $k$ is an integer such that $|k|\leq N-1$ and $ N-1+k$ being
even, and  yields  two different values $M^{(\pm)}$ for $M$ in
(\ref{fusion-5}), corresponding respectively  to
$H^{(\pm)}_i(u|\{\e^{(\g)}_l\})$, \bea
M^{(\pm)}=\frac{1}{2}\lt(N-1\mp k\rt). \label{Eign-fuc} \eea The
proof of the above constraint is relegated to Appendix B.

Finally, we have that if the boundary parameters $\{\a^{(\g)}_l\}$
satisfy any of the constraints (\ref{Restr-1}), the eigenvalues of
the fundamental transfer matrix $t(u)$ are given by  \bea
\Lambda^{(\pm)}(u)&=& H^{(\pm)}_1(u|\{\e^{(\g)}_l\})
\frac{Q^{(\pm)}(u+\eta)}{Q^{(\pm)}(u)}+
H^{(\pm)}_2(u|\{\e^{(\g)}_l\})
\frac{Q^{(\pm)}(u-\eta)}{Q^{(\pm)}(u)},\label{result}\eea where $
Q^{(\pm)}(u)= \prod_{j=1}^{M^{(\pm)}}
\s(u-u^{(\pm)}_j)\s(u+u^{(\pm)}_j+\eta)$ and the parameters
$\{u^{(\pm)}_j\}$ respectively satisfy the Bethe Ansatz equations,
\bea &&\frac{H^{(\pm)}_{2}(u^{(\pm)}_j|\{\e^{(\g)}_l\})}
{H^{(\pm)}_{1}(u^{(\pm)}_j|\{\e^{(\g)}_l\})}
=-\frac{Q^{(\pm)}(u^{(\pm)}_j+\eta)}{Q^{(\pm)}(u^{(\pm)}_j-\eta)},
\qquad j=1,\ldots,M^{(\pm)}.\label{BAE-2}\eea Indeed one can
verify that both $\Lambda^{(\pm)}(u)$ given by (\ref{result}) have
the desirable properties (\ref{Eigen-Anal})-(\ref{Eigen-semi})
provided that the constraint (\ref{Restr-1}) and the Bethe Ansatz
equations (\ref{BAE-2}) are satisfied.  In the trigonometric limit
$\tau\rightarrow +i\infty$, we recover the results in \cite{Yan06}
(for details see Appendix C). The completeness \cite{Nep03,Yan06}
of the eigenvalues $\{\Lambda^{(\pm)}(u)\}$ in the trigonometric
case suggests that for a given set of bulk and boundary parameters
satisfying the constraint (\ref{Restr-1}), the eigenvalues
$\Lambda^{(-)}(u)$ and $\Lambda^{(+)}(u)$ {\it together}
constitute the complete set of eigenvalues of the transfer matrix
$t(u)$ of the open XYZ spin chain.

Therefore the complete set of the energy eigenvalues $E^{(\pm)}$
of the Hamiltonian (\ref{Ham}), when the boundary parameters
satisfy the constraint (\ref{Restr-1}), are  respectively given by
\bea E^{(\pm)}\hspace{-0.1truecm}=\hspace{-0.1truecm}
\frac{\s(\eta)}{\s'(0)}\lt\{2\sum_{j=1}^{M^{(\pm)}}\lt(
\zeta(u^{(\pm)}_j)-\zeta(u^{(\pm)}_j+
\eta)\rt)\hspace{-0.1truecm}+\hspace{-0.1truecm}(N-1)\zeta(\eta)
+\sum_{\g=\pm}\sum_{l=1}^3\zeta(\mp\e^{(\g)}_l\a^{(\g)}_l)
\rt\},\label{H-Eign}\eea where the parameters $\{u^{(\pm)}_j\}$
respectively satisfy the Bethe Ansatz equations (\ref{BAE-2}).

Some remarks are in order. Firstly, in \cite{Fan96} a generalized
algebraic Bethe Ansatz \cite{Bax73} was used to diagonalize the
open XYZ spin chain when boundary parameters satisfying a
constraint. However, only {\it partial\/} eigenvalues of transfer
matrix corresponding to our $\Lambda^{(+)}(u)$ were obtained there
and it is not clear whether the approach in \cite{Fan96} can give
all the eigenvalues of the transfer matrix. (See also
\cite{Cao03}, where partial eigenvalues of the transfer matrix of
the open XXZ spin chain were obtained.) Indeed, it is found in
this paper that both $\L^{(\pm)}(u)$ are needed to constitute a
complete set of eigenvalues. This implies that two sets of Bethe
Ansatz equation (\ref{BAE-2}) and therefore presumably two
pseudo-vacua are required. Unfortunately, it is not yet clear how
to construct the second pseudo-vacuum in the generalized algebraic
Bethe Ansatz approach. Secondly, when the crossing parameter
$\eta$ is equal to $\frac{1}{p+1}$ with $p$ being a non-negative
integer (i.e. roots of unity case), the corresponding fusion
hierarchy truncates as in the trigonometric case \cite{Baz96}, and
one may generalize the method developed in \cite{Nep04} to obtain
the corresponding functional relations obeyed by the transfer
matrices and therefore the eigenvalues of the fundamental transfer
matrix $t(u)$. If the boundary parameters satisfy the constraint
(\ref{Restr-1}), the complete set of eigenvalues and the
associated Bethe Ansatz equations, in the roots of unity case, are
expected to be {\it still\/} given by (\ref{result}) and
(\ref{BAE-2}) respectively. Thirdly, it would be interesting to
determine the conditions for which the limit (\ref{Qbar}) exists.
We have seen that (\ref{result})-(\ref{BAE-2}) solve the open XYZ
chain with generic values of $\eta$ if the constraint
(\ref{Restr-1}) is satisfied. This suggests that, for generic
values of $\eta$, the constraint (\ref{Restr-1}) may be a
necessary condition for the existence of the limit (\ref{Qbar}).
It would  be interesting to explicitly evaluate the $Q$-operator
directly from Eq. (\ref{Qbar}).


\section{Conclusions}
\label{Con} \setcounter{equation}{0}

We have argued that the Baxter's $Q$-operator for the
 open XYZ chain is given by the $j \rightarrow \infty$ limit of the
transfer matrix with spin-$j$ auxiliary space. This together with
the fusion hierarchy leads to the Baxter's $T$-$Q$ relations
(\ref{T-Q}) and (\ref{T-Q-1}). These $T$-$Q$ relations, together
with the additional properties
(\ref{Eigen-Anal})-(\ref{Eigen-semi}), enable us to successfully
determine the eigenvalues (\ref{result}) of the fundamental
transfer matrix and the associated Bethe Ansatz equations
(\ref{BAE-2}). For a given set of bulk and boundary parameters
satisfying the constraint (\ref{Restr-1}), the eigenvalues
$\Lambda^{(-)}(u)$ and $\Lambda^{(+)}(u)$ {\it together}
constitute the complete set of eigenvalues of the fundamental
transfer matrix $t(u)$, which leads to the complete spectrum
(\ref{H-Eign}) of the Hamiltonian (\ref{Ham}). Our derivation for
the $Q$-operator and associated $T$-$Q$ relation can be applied to
other models which share $sl_2$-like fusion rule. Moreover, it
would be interesting to generalize our method to integrable models
associated with  higher rank algebras \cite{Yan04,Yan05,Doi03}.

\section*{Acknowledgments}
We thank Ryu Sasaki for critical comments and useful discussions.
Financial support from  the Australian Research Council is
gratefully acknowledged.


\section*{Appendix A: Determinants and crossing relations}
\setcounter{equation}{0}
\renewcommand{\theequation}{A.\arabic{equation}}
In this appendix, we compute the determinants (\ref{Delta-1}) and
(\ref{Delta-2}) of the K-matrices. In so doing, we will also prove
the crossing relations (\ref{K-cross-3}) and (\ref{K-cross-4}).

One rewrites $K^-(u)$ in (\ref{K-matrix1}) as \bea
K^-(u)=\sum_{\a=0}^3f_{\a}(u)\s^{\a},\quad \s^0= I,\,\s^1=\s^x,\,
\s^2=\s^y,\,\s^3=\s^z.\no\eea It is easy checked that the
coefficient function $\{f_{\a}(u)\}$ satisfy \bea f_0(-u)=f_0(u),
\qquad f_i(-u)=-f_i(u),\quad {\rm for}\,\, i=1,2,3.\no\eea The
above parity relations lead to the following unitarity relation of
$K^-(u)$ \bea K^-(u)K^-(-u)=\Delta_2(u)\times {\rm id},\quad
\Delta_2(u)=\sum_{\a=0}^3f_{\a}(u)f_{\a}(-u).\label{A-1} \eea From
the expression (\ref{K-matrix1}) of $K^-(u)$ and after direct
calculation, we find that \bea
\Delta_2(u)=\prod_{l=1}^3\frac{\s(\a^{(-)}_l+u)\s(\a^{(-)}_l-u)}
{\s(\a^{(-)}_l)\s(\a^{(-)}_l)}.\label{D-fuction-2}\eea

We now follow a method similar to that developed in \cite{Hou98}.
Noting $R_{1,2}(-\eta)=-2P^{(-)}_{12}$ and PT-symmetry (\ref{PT})
of the R-matrix, the reflection equation (\ref{RE-V}), when
$u_1=u,\,u_2=u+\eta$, reads \bea
P^{(-)}_{12}K^-_1(u)R_{1,2}(2u+\eta)K^-_2(u+\eta)=K^-_2(u+\eta)
R_{1,2}(2u+\eta)K^-_1(u)P^{(-)}_{12}.\no\eea Since the dimension
of the image of $P^{(-)}_{12}$ is equal to one, the above equation
becomes \bea K^-_2(u+\eta) R_{1,2}(2u+\eta)K^-_1(u)P^{(-)}_{12}&=&
P^{(-)}_{12}K^-_1(u)R_{1,2}(2u+\eta)K^-_2(u+\eta)P^{(-)}_{12}\no\\
&=&tr_{12}\lt(P^{(-)}_{12}K^-_1(u)R_{1,2}(2u+\eta)K^-_2(u+\eta)\rt)
\,P^{(-)}_{12}\no\\
&\stackrel{(\ref{Delta-1})}{=}&{\rm
Det}\{K^-(u+\eta)\}\,P^{(-)}_{12}. \label{A-3}\eea It is
well-known that the completely anti-symmetric vector with unit
normal in $\Cb^2\otimes\Cb^2$ is \bea
\frac{1}{\sqrt{2}}\lt(|1\rangle\otimes|2\rangle-|2\rangle\otimes|1\rangle\rt)
=\lt({\rm id}\otimes
V\rt)\lt(\frac{1}{\sqrt{2}}\sum_{i=1}^2|i\rangle\otimes|i\rangle\rt),
\label{A-4}\eea where $\{|1\rangle, |2\rangle\}$ are the
orthnormal basis of $\Cb^2$. Acting both sides of (\ref{A-3}) on
the vector (\ref{A-4}) and after straightforward calculating, we
find that (\ref{A-3}) is indeed equivalent to \bea
\bar{K}^-(u)={\rm
Det}\{K^-(u+\eta)\}\lt\{VK^-(u+\eta)V\rt\}^{-1}.\no\eea Using
(\ref{A-1}), we have \bea \bar{K}^-(u)=\frac{{\rm
Det}\{K^-(u+\eta)\}}{\Delta_2(u+\eta)}\,VK^-(-u-\eta)V.
\label{A-5}\eea Taking trace of both sides of (\ref{A-5}), we
have, for the L.H.S. of the resulting relation, \bea {\rm L.H.S.}
&=&tr\lt(\bar{K}^-(u)\rt)\stackrel{(\ref{K-cross-1})}{=}tr_{12}\lt(
P_{12}R_{1,2}(-2u-2\eta)\{K^-_2(u)\}^{t_2}\rt)\no\\
&=&tr_{2}\lt(tr_1\lt\{
P_{12}R_{1,2}(-2u-2\eta)\rt\}\{K^-_2(u)\}^{t_2}\rt)\no\\
&=&\lt(a(-2u-2\eta)+c(-2u-2\eta)\rt)tr_2\lt(K^-_2(u)\rt)\no\\
&=&\lt(a(-2u-2\eta)+c(-2u-2\eta)\rt)\frac{\s(2u)}{\s(u)}\no\\
&=&-\frac{\s(2u)\s(2u+2\eta)}{\s(\eta)\s(u+\eta)}.\label{B-3}\eea
In the above deriving, we have used the following relations: \bea
tr_1\lt(P_{12}R_{1,2}(u)\rt)&=&\lt(a(u)+c(u)\rt)\times {\rm
id},\no\\
a(-2u-2\eta)+c(-2u-2\eta)&=&-\frac{\s(2u+2\eta)\s(u)}
{\s(\eta)\s(u+\eta)},\no\eea which can be obtained from
(\ref{identity})-(\ref{r-func}), and  the following identities of
the elliptic functions: \bea \theta
\lt[\begin{array}{c}\frac{1}{2}\\[2pt]\frac{1}{2}\end{array}\rt]
(2u,2\tau)&=&\theta
\lt[\begin{array}{c}\frac{1}{2}\\[2pt]\frac{1}{2}\end{array}\rt]
(\tau,2\tau)\,\times\,\frac{\s(u)\s(u+\frac{1}{2})}
{\s(\frac{\tau}{2})\s(\frac{1}{2}+\frac{\tau}{2})},\no\\
\theta \lt[\begin{array}{c}0\\\frac{1}{2}\end{array}\rt]
(2u,2\tau)&=&\theta
\lt[\begin{array}{c}0\\\frac{1}{2}\end{array}\rt]
(0,2\tau)\,\times\,\frac{\s(u-\frac{\tau}{2})\s(u+\frac{1}{2}+\frac{\tau}{2})}
{\s(-\frac{\tau}{2})\s(\frac{1}{2}+\frac{\tau}{2})}.\no\eea On the
other hand, for the R.H.S. of the resulting relation, we obtain
\bea {\rm R.H.S.} &=&
\frac{{\rm Det}\{K^-(u+\eta)\}}{\Delta_2(u+\eta)}tr\lt(VK^-(-u-\eta)V\rt)\no\\
&=&\frac{{\rm Det}\{K^-(u+\eta)\}}{\Delta_2(u+\eta)}tr\lt(V^2K^-(-u-\eta)\rt)\no\\
&=&-\frac{{\rm Det}\{K^-(u+\eta)\}}{\Delta_2(u+\eta)}tr\lt(K^-(-u-\eta)\rt)\no\\
&=&-\frac{{\rm Det}\{K^-(u+\eta)\}}{\Delta_2(u+\eta)}\,
\frac{\s(2u+2\eta)}{\s(u+\eta)}. \label{B-4}\eea Comparing with
(\ref{B-3}) and (\ref{B-4}), we have \bea \frac{{\rm
Det}\{K^-(u+\eta)\}}{\Delta_2(u+\eta)}=\frac{\s(2u)}{\s(\eta)}.
\label{B-5}\eea Now (\ref{D-2-1}) follows from (\ref{D-fuction-2})
and (\ref{B-5}). Similarly one can prove (\ref{D-3}) using
(\ref{K-matrix2}). Finally, (\ref{A-5}) and (\ref{B-5}) give rise
to (\ref{K-cross-3}), and (\ref{K-cross-4}) can be similarly be
proven from (\ref{K-matrix2}).

\section*{Appendix B: Constraint condition (\ref{Restr-1})}
\setcounter{equation}{0}
\renewcommand{\theequation}{B.\arabic{equation}}
One can easily check that both forms (\ref{T-Q-3}) of eigenvalues
of $t(u)$ satisfy the first quasi-periodicity of (\ref{Eign-Per})
as required. Now we investigate  the second quasi-periodic
property. The definitions (\ref{H-1}) and (\ref{H-2}) of
$\{H^{(\pm)}_i(u|\{\e^{(\g)}_l\})\}$ imply \bea
&&\hspace{-1.42truecm}H^{(\pm)}_1(u+\tau|\{\e^{(\g)}_l\})
=e^{-2i\pi\lt((N+3)(2u+\tau)+4\eta\rt)}
e^{-2i\pi\lt(\pm\sum_{\g=\pm}\sum_{l=1}^3\e^{(\g)}_l\a^{(\g)}_l\rt)}
H^{(\pm)}_1(u|\{\e^{(\g)}_l\}),\no\\
&&\hspace{-1.42truecm}H^{(\pm)}_2(u+\tau|\{\e^{(\g)}_l\})
=e^{-2i\pi\lt((N+3)(2u+\tau)+(2N+2)\eta\rt)}
e^{-2i\pi\lt(\mp\sum_{\g=\pm}\sum_{l=1}^3\e^{(\g)}_l\a^{(\g)}_l\rt)}
H^{(\pm)}_2(u|\{\e^{(\g)}_l\}).\no\eea Keeping the above
quasi-periodic properties and the definition (\ref{fusion-5}) of
$Q(u)$  in mind, in order that (\ref{T-Q-3}) have the second
property of (\ref{Eign-Per}) one needs that the boundary
parameters $\{\a^{(\g)}_l\}$ satisfy the following constraint \bea
\sum_{\g=\pm}\sum_{l=1}^3\e^{(\g)}_l\a^{(\g)}_l=k\eta\quad{\rm
mod}(1),\label{C-1}\eea where $k$ is an integer such that $|k|\leq
N-1$ and $ N-1+k$ being even. For such  integer $k$, let us
introduce two non-negative integers $M^{(\pm)}$:
$M^{(\pm)}=\frac{1}{2}\lt(N-1\mp k\rt)$.  Moreover, associated
respectively with $H^{(\pm)}_i(u|\{\e^{(\g)}_l\})$, one introduces
$Q^{(\pm)}(u)$,\bea Q^{(\pm)}(u)= \prod_{j=1}^{M^{(\pm)}}
\s(u-u^{(\pm)}_j)\s(u+u^{(\pm)}_j+\eta).\no\eea Hence the
eigenvalues of the transfer matrix take the following forms: \bea
H^{(\pm)}_1(u|\{\e^{(\g)}_l\})\,\frac{Q^{(\pm)}(u+\eta)}{Q^{(\pm)}(u)}
+H^{(\pm)}_2(u|\{\e^{(\g)}_l\})\,\frac{Q^{(\pm)}(u-\eta)}{Q^{(\pm)}(u)}.
\label{C-2}\eea At this stage, we still have freedom to chose the
discrete parameters $\{\e^{(\g)}_l\}$. Now we apply the
semi-classical property (\ref{Eigen-semi}) of $\L(u)$ to obtain a
restriction to this freedom.

For this purpose, let us introduce $\L^{(\pm)}(u)$,  corresponding
to (\ref{C-2}), \bea \L^{(\pm)}(u)=
H^{(\pm)}_1(u|\{\e^{(\g)}_l\})\,\frac{Q^{(\pm)}(u+\eta)}{Q^{(\pm)}(u)}
+H^{(\pm)}_2(u|\{\e^{(\g)}_l\})\,\frac{Q^{(\pm)}(u-\eta)}{Q^{(\pm)}(u)},
\label{C-3}\eea  and define
$\bar{\a}^{(\g)}_l=\lim_{\eta\rightarrow 0}\a^{(\g)}_l$. Since the
value of the discrete parameters $\{\e^{(\g)}_l\}$ does not change
when taking the limit of $\eta\rightarrow 0$, the constraint
(\ref{C-1}) in this limit becomes \bea
\sum_{\g=\pm}\sum_{l=1}^3\e^{(\g)}_l\bar{\a}^{(\g)}_l=0\quad{\rm
mod}(1).\label{C-4}\eea Substituting (\ref{C-3}) into the
semi-classical property (\ref{Eigen-semi}) of the eigenvalues of
the fundamental transfer matrix $t(u)$, we have for the L.H.S. of
the resulting relation, \bea {\rm L.H.S.}&=& \lim_{\eta\rightarrow
0}\lt\{\s^{2N}(\eta)\L^{(\pm)}(u)\rt\}\no\\
&=&\s^{2N}(u)\lt\{
\prod_{\g=\pm}\prod_{l=1}^3\frac{\s(u\pm\e^{(\g)}_l\bar{\a}^{(\g)}_l)}
{\s(\e^{(\g)}_l\bar{\a}^{(\g)}_l)}+
\prod_{\g=\pm}\prod_{l=1}^3\frac{\s(u\mp\e^{(\g)}_l\bar{\a}^{(\g)}_l)}
{\s(\e^{(\g)}_l\bar{\a}^{(\g)}_l)}\rt\}.\no\eea Comparing both
side of the resulting relation, (\ref{Eigen-semi}) is equivalent
to the following relation \bea
\prod_{\g=\pm}\prod_{l=1}^3\frac{\s(u\pm\e^{(\g)}_l\bar{\a}^{(\g)}_l)}
{\s(\e^{(\g)}_l\bar{\a}^{(\g)}_l)}+
\prod_{\g=\pm}\prod_{l=1}^3\frac{\s(u\mp\e^{(\g)}_l\bar{\a}^{(\g)}_l)}
{\s(\e^{(\g)}_l\bar{\a}^{(\g)}_l)}=\lim_{\eta\rightarrow
0}\hspace{-0.1cm}\lt\{tr(K^+(u)K^-(u))\rt\}.\label{C-6}\eea
Keeping the constraint (\ref{C-4}) among $\{\bar{\a}^{(\g)}_l\}$,
after straightforward calculation, we find that (\ref{C-6}) is
satisfied if and only if the following constraint among the
discrete parameters $\{\e^{(\g)}_l\}$ were fulfilled \bea
\prod_{\g=\pm}\prod_{l=1}^3\e^{(\g)}_l=-1. \label{C-7}\eea
(\ref{C-1}) and (\ref{C-7}) are nothing but the constraints listed
in  (\ref{Restr-1}).

\section*{Appendix C: Trigonometric limit}
\setcounter{equation}{0}
\renewcommand{\theequation}{C.\arabic{equation}}
In this Appendix, we consider the trigonometric limit
$\tau\rightarrow +i\infty$ of our result. The definition of the
elliptic functions (\ref{Function-a-b})-(\ref{Function}) imply
\bea \s(u+\frac{\tau}{2})=e^{-i\pi(u+\frac{1}{2}+\frac{\tau}{4})}
\,\,\theta\lt[\begin{array}{c}0\\\frac{1}{2}\end{array}\rt](u,\tau),
\label{AD-1}\eea and the following asymptotic behaviors \bea
&&\lim_{\tau\rightarrow
+i\infty}\s(u)=-2e^{\frac{i\pi\tau}{4}}\sin \pi
u+\ldots,\label{AD-2}\\
&&\lim_{\tau\rightarrow
+i\infty}\theta\lt[\begin{array}{c}0\\\frac{1}{2}\end{array}\rt](u,\tau)=
1+\ldots.\label{AD-3}\eea The above asymptotic behaviors lead to
the well-known result: \bea \lim_{\tau\rightarrow
+i\infty}R(u)=\frac{1}{\sin\,\pi\eta}\lt(\begin{array}{cccc}\sin\pi(u+\eta)
&&&\\&\sin\,\pi u&\sin\,\pi\eta&\\
&\sin\,\pi\eta&\sin\,\pi u&\\&&&\sin\pi(u+\eta)\end{array}\rt).
\label{t-r-matrix}\eea After reparameterizing the boundary
parameters \bea \a_1^{(\mp)}=\pm\a_{\mp},\quad
\a^{(\mp)}_2=\mp\b_{\mp}+\frac{1}{2},
\quad\a^{(\mp)}_3=-\theta_{\mp}+\frac{1}{2}+\frac{\tau}{2},\label{reparam}\eea
and using (\ref{AD-1})-(\ref{AD-3}), one has \bea
&&\hspace{-1.8truecm}\lim_{\tau\rightarrow +i\infty}K^-(u)\equiv
K_{(t)}^-(u)=\frac{1}{2\sin\pi\a_-\cos\pi\b_-}\lt\{2\sin\pi\a_-\cos\pi\b_-\cos
\pi u\,I\rt.\no\\
&&\quad +\lt.2i\cos\pi\a_-\sin\pi\b_-\sin \pi u\,\s^z+\sin 2\pi
u\lt(\cos\pi\theta_-\,\s^x-\sin\pi\theta_-\,\s^y\rt)\rt\},\label{t-K-matrix-1}\\
&&\hspace{-1.8truecm}\lim_{\tau\rightarrow +i\infty}K^+(u)\equiv
K_{(t)}^+(u)=\lt.K_{(t)}^-(-u-\eta)\rt|_{(\a_-,\b_-,\theta_-)\rightarrow
(-\a_+,-\b_+,\theta_+)}.\label{t-K-matrix-2}\eea Hence our
R-matrix and K-matrices, when taking the trigonometric limit,
reduce to the corresponding trigonometric ones used in
\cite{Yan06} after a rescaling of the corresponding parameters. In
order for the constraint (\ref{Restr-1}) to be still satisfied
after the limit, one needs further to require that
$\e^{(-)}_3=-\e^{(+)}_3$. Then, the resulting constraint becomes
that in \cite{Yan06}. Moreover, our
$\{H^{(\pm)}_i(u|\{\e^{(\g)}_l\})\}$ in the limit give rise to the
corresponding trigonometric ones in \cite{Yan06} up to some
constants (due to the constant factors appearing in
(\ref{t-r-matrix}), (\ref{t-K-matrix-1})-(\ref{t-K-matrix-2})).
Therefore, our result recovers that of \cite{Yan06} in the
trigonometric limit.


\end{document}